\pdfoutput=1
\documentclass{rnaastex}

\begin{document}

\title{On the Nova Rate in M87}

\email{ashafter@sdsu.edu, akundu@eurekasci.com, mhenze@sdsu.edu}

\author{A. W. Shafter}
\altaffiliation{Department of Astronomy, San Diego State University}

\author{A. Kundu}
\altaffiliation{Eureka Scientific Inc.}

\author{M. Henze}
\altaffiliation{Department of Astronomy, San Diego State University}





\section{} 
The question of whether the luminosity-specific nova rate (LSNR)
varies across differing Hubble-type galaxies is poorly understood.
A particularly important study of novae in M87 was undertaken recently
by \citet{shr16}, who
searched for novae in a series of HST images
originally acquired in {\it HST\/} program \#10543 (PI: Baltz).
Based on their discovery of 32 nova candidates
\citet{shr16} estimated that the nova rate in M87 is
$363_{-45}^{+33}$ yr$^{-1}$, which is more than double
the rates found in previous ground-based studies \citep[e.g.,][]{cur15, sha00}.
In an attempt to reconcile the difference between
Shara et al.'s result and previous nova rate estimates,
we have undertaken an independent analysis of the {\it HST\/} data.

We searched for novae by
using an automated algorithm that independently detected variable
sources against the rapidly varying, local background of M87.
Candidates required at least two epochs of $S/N > 2$ within a fixed aperture.
All objects were confirmed through blinking, and were classified manually
via examination of their light curves.
A total of 33 transient sources were identified, and
their light curves are shown in Figure 1. Based on a qualitative examination
of the data,
we concluded that 16 sources are
almost certainly classical novae
(Group A: rapid rise to maximum light followed by a slower decline).
Another five have incomplete lightcurves, but are nevertheless likely to be
novae (Group B: maximum light covered with partial decline).The
remaining 12 sources are possibly novae, but the lightcurves
are ambiguous (Group C: maximum missed, declining light curve).
Most, but not all, of the candidates shown in Figure~1
are objects identified as novae by \citet{shr16}.
A definitive determination that the sources are bona fide
novae requires either spectroscopy, for which there are no data,
or reliable and complete nova light curves, which are only available for
roughly two-thirds of the nova candidates.
Thus, we cannot completely rule out the possibility that as many as
one-third of the candidates could be sources other than novae.
The symmetric light curve of nova candidate \#14, for example, suggests
that this object could be a microlensing event.

Following the Monte Carlo procedure outlined in \citet{cur15},
we determined several estimates of the M87 nova rate
under a variety of assumptions.
We modeled the nova luminosity function using
the peak magnitudes and decline rates for group A novae. The
detection completeness was determined via a blind search for
artificially generated nova-like features within a range of brightnesses.
We find a rate of $139_{-14}^{+16}~\mathrm{yr}^{-1}$
{\it in the $202''\times202''$ surveyed region\/} of M87
when all 33 nova candidates are
included.
If we restrict the sample to only Groups A and B
(transients thought to be very likely novae based on their
light curve behavior), the rate drops to $88_{-11}^{+14}~\mathrm{yr}^{-1}$.
When only the strongest nova candidates are included,
the nova rate in the
surveyed region decreases  to $68_{-9}^{+10}$~yr$^{-1}$. The corresponding
LSNRs ($K_{\mathrm 2MASS}=5.97$ in the surveyed region) are 
$6.4\pm1.5$, $4.1\pm1.0$, and $3.2\pm0.8$, novae per year per
$10^{10} L_{K,\odot}$, respectively.

Estimating the total nova rate in M87 rests on
an uncertain extrapolation from
the surveyed region to the entire galaxy.
Extragalactic nova studies usually trace the stellar mass
via the infrared $K$-band light, which is easier to
measure. However, M87 is surrounded by intracluster stars and gas, which
makes it extremely difficult to
accurately measure M87's surface brightness in the outskirts of
the galaxy.

To explore the effect on the global nova rate, we have
tested two different extrapolations: one based on
the photometry of \citet{kor09} used by \citet{shr16}, and another
based on the photometry of \citet{coh86} used in the ground-based
studies of \citet{sha00} and \citet{cur15}.
We find significantly higher nova rates for the \citet{kor09}
extrapolations (which include the extended halo of M87):
$339_{-33}^{+37}$, $215_{-27}^{+35}$, and
$166_{-23}^{+25}$ per year,
compared with the \citet{coh86} extrapolations:
$201_{-20}^{+22}$, $128_{-16}^{+21}$, and $99_{-15}^{+16}$, per year
for groups A+B+C, A+B, and A, respectively.

\section{}

Another source of uncertainty is whether the LSNR varies
with galactocentric radius, which remains unaddressed.
In M87, it is possible that
the LSNR might be enhanced in the inner regions of the galaxy
as a result of past merger events or globular cluster disruptions
that have increased the number of nova progenitor binaries there.
In fact, an earlier {\it HST}/STIS study by \citet{mad07}
found 11 out of 13 ultraviolet
transients within just 9$''$ of the nucleus.
An extrapolation to the entire galaxy
would suggest a M87 nova rate of $\sim$1000~yr$^{-1}$.
Such a high rate is clearly ruled
out by wide-field ground-based observations \citep{cur15},
providing evidence that the nova rate is enhanced near the galaxy center.

\begin{figure}
\includegraphics[scale=0.55,angle=0]{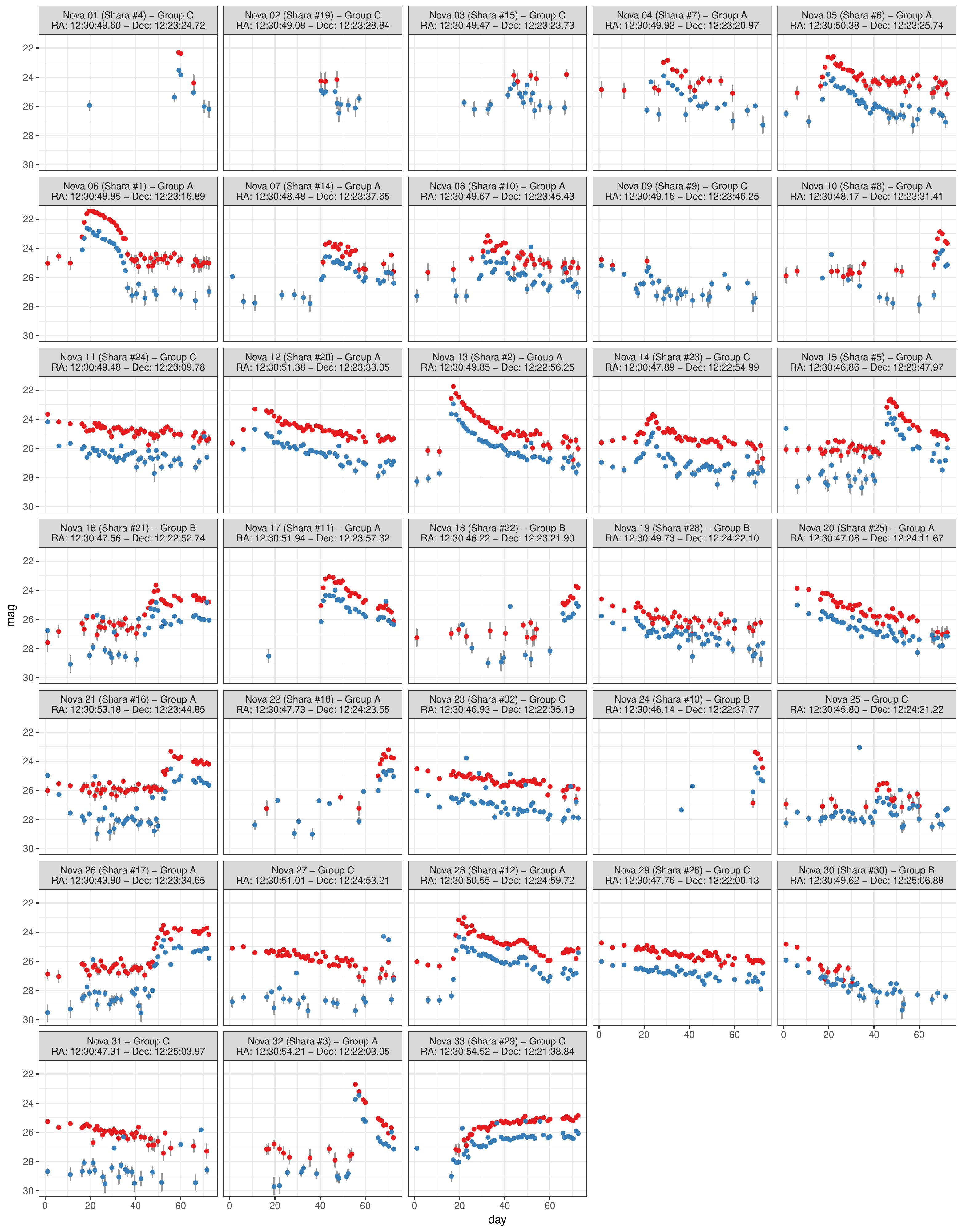}
\caption{The light curves (Red: F814W; Blue: F660W, offset by $+1$ mag)
for the 33 nova candidates ordered by distance from the center of M87.
The coordinates of each nova candidate are given, along with our
judgement about the reliability of the nova identification (Groups
A, B, C; see text for details).}
\end{figure}

In conclusion, our results are in general agreement with those of \citet{shr16},
although we argue that their rate of 363~yr$^{-1}$ is likely an upper limit,
and that the global nova rate in M87 remains
uncertain, both due to the difficulty in identifying bona fide novae
from incomplete lightcurves,
and to extrapolating observations near the center of M87 to the
entire galaxy.


A.W.S. and A.K. acknowledge support from HST grant AR-12139.



\begin{thebibliography}{}

\bibitem[Curtin et al.(2015)]{cur15} Curtin, C., Shafter, A. W., Pritchet, C. J., et al.\ 2015, ApJ, 811, 34

\bibitem[Cohen(1986)]{coh86} Cohen, J. G.\ 1986, \aj, 92, 1039

\bibitem[Kormendy et al.(2009)]{kor09} Kormendy, J., Fisher, D. B., Cornell, M. E., et al.\ 2009, ApJS, 182, 216

\bibitem[Madrid et al.(2007)]{mad07} Madrid, J.~P., Sparks,
W.~B., Ferguson, et al.\ 2007, \apjl, 654, L41

\bibitem[Shafter et al.(2000)]{sha00} Shafter, A.~W.,
Ciardullo, R., \& Pritchet, C.~J.\ 2000, \apj, 530, 193

\bibitem[Shara et al.(2016)]{shr16} Shara, M. M., Doyle, T. F., Lauer, T. R., et al. 2016, ApJS, 227, 1

\end{thebibliography}
\end{document}